\documentstyle[12pt]{article} 
\newlength{\dinwidth}
\newlength{\dinmargin}
\newcommand{\tr}{\mbox{tr}}

\newcommand{\NP}[1]{{\em Nucl.\ Phys.\ }{\bf #1}}
\newcommand{\ZP}[1]{{\em Z.\ Phys.\ }{\bf #1}}

\newcommand{\PL}[1]{{\em Phys.\ Lett.\ }{\bf #1}}

\newcommand{\PR}[1]{{\em Phys.\ Rev.\ }{\bf #1}}

\begin{document}
\input epsf
\pagestyle{empty}
\renewcommand{\baselinestretch}{1}

\vspace*{2.8cm}
\begin{center}
{ \bf STRONG WW SCATTERING
 AND RESONANCES AT LHC}

\vspace{1.5cm}

{J. R. Pel\'aez\\ Departamento de F\'{\i}sica Te\'orica.\\
Universidad Complutense. \\28040 Madrid. Spain}
\end{center}


\vskip .2cm 

\begin{abstract}
The low energy  dynamics of the general strongly interacting symmetry
breaking sector can be easily described using effective chiral Lagrangians.
Indeed, the enhancement in WW scattering at
LHC, that would imply the existence of such an strong interaction, can be  
described with just two chiral parameters.
These techniques have been shown to reproduce
remarkably well the low-energy pion-pion scattering data, which follows
a similar formalism.
In this work we first review the LHC sensitivity to those two chiral parameters
(in the hardest case of non-resonant low-energy WW scattering).
Later it is shown how we can predict the general resonance spectrum 
of the strongly interacting symmetry breaking sector. For that purpose, we
 use the inverse amplitude method which is also very successful 
reproducing the lightest hadronic resonances from
data in the low-energy non resonant region.
We thus present an study of the regions in parameter space
where one, two or no resonances may appear.
\end{abstract}
\begin{center}
Contribution to the XXXIInd Rencontres de Moriond:

 "Electroweak Interactions and Unified Theories"
\end{center}

\newpage

\renewcommand{\baselinestretch}{1.5}

{\bf The Strongly Interacting Symmetry Breaking Sector}

\vskip .2 cm

In the Standard Model we need a Symmetry Breaking Sector (SBS) in order
to explain the masses of the electroweak gauge bosons. Otherwise, the interactions
of these particles would not be renormalizable and would violate unitarity.
This fact is specially evident in the scattering of longitudinal gauge bosons ($V_L$).

The simplest model of $SU(2)_L\times U(1)_Y$ spontaneous breaking, preserves
renormalizability and restores unitarity by adding a complex doublet to the Standard Model
(SM).
Three of these new degrees of freedom are nothing but the Goldstone Bosons (GB) that
become the longitudinal components of the gauge bosons through the Higgs mechanism.
The remaining scalar field, known as the Higgs boson, should be observable. 
That is the Minimal Standard Model (MSM).  But this is not
the only way to build the SBS. Indeed there are other models with many more particles like
the Higgs, no Higgs at all,  vector fields, etc... whose masses are expected in the range of a 
few TeV or less.

If we do not find at LHC particles much lighter than 1 TeV 
belonging to the SBS, then the interactions of
longitudinal gauge bosons would grow until they 
become strong. In such case, we expect an enhancement
in the process $V_LV_L\longrightarrow V_LV_L$ at LHC. Another  typical 
feature of strong interactions that saturate unitarity are resonances, and
we also expect that some of them would show up at LHC.
The above two sentences about the strongly interacting SBS
may seem too vague. In addition, strong models
are non-perturbative and it is pretty hard to obtain reliable predictions 
on observables or on the possible resonances. That is why here we will
address the following two questions:

\begin{center}
What could we measure?

What resonance spectrum do we expect?

\end{center}
\vskip .1 cm

{\bf The  Electroweak Effective Chiral Lagrangian}

\vskip .2 cm

The underlying theory that breaks 
the SM $SU(2)_L\times U(1)_Y$ group down to $U(1)_{EM}$ is
unknown to a large extent. 
Basically, what we know  is the following:

$\bullet$ There should be a system with a {\em global} symmetry breaking
yielding three GB. 

$\bullet$ The scale 
of this new interactions is $v\simeq250\mbox{GeV}$. 

$\bullet$ The electroweak $\rho$ parameter is very close to one.

\noindent
This last requirement is most naturally satisfied if the 
electroweak SBS respects the so called 
custodial symmetry
$SU(2)_{L+R}$ $^{1})$. 
Demanding just three GB, we are 
lead to a breaking from $SU(2)_L\times SU(2)_R$ down to
$SU(2)_{L+R}$ $^{2,3})$. Most of the models of symmetry
breaking, including the Minimal Standard Model follow this breaking scheme
. Formally it is the same breaking pattern of the chiral symmetry 
in QCD  with just two massless quarks. 
Although a rescaled version
of QCD is not valid in our case, we can still borrow 
the formalism $^4)$, which works remarkably well with the 
pion-pion scattering data $^5)$.

In our case we are interested in the longitudinal gauge bosons,
which somehow are {\em equivalent} to the GB. 
Hence, the chiral lagrangian is built as a derivative 
expansion using GB fields. In the amplitudes, the derivatives
become external momenta or energy. It is therefore a low-energy expansion.
There is only one possible term with two derivatives
that respects the above symmetry breaking pattern:
\begin{equation}
{\cal L}^{(2)}=\frac{v^2}{4}\tr D_\mu UD^\mu U^\dagger
\label{NLSM}
\end{equation}
where the GB fields $\pi^i$ are
collected in the $SU(2)$ matrix
$U=\exp(i\pi^i \sigma^i/v)$ and $D_\mu$ is 
the usual $SU(2)_L\times U(1)_Y$ covariant derivative. 
It is important to remark that the above lagrangian
only depends on the symmetry 
structure and the scale. Its predictions for $V_LV_L$ scattering
are universal. The dependence on the different models appears 
at next order through
two phenomenological parameters: 
\begin{equation}
{\cal L}^{(4)}=
L_1 \left( \tr D_\mu UD^\mu U^\dagger \right)^2
+ L_2 \left( \tr D_\mu UD^\nu U^\dagger \right)^2
\label{L4}
\end{equation}
Notice that we have only given the operators which are relevant
for $V_LV_L\rightarrow V_LV_L$ (working also at lowest order
in the electroweak corrections). Nevertheless, we have to keep in mind that this formalism is only valid
when there are no {\em light} ($\simeq$ few hundred GeV) particles.
Otherwise we should have to include such states in our description.

 The values of $L_1$ and $L_2$ depend on the model, but we
expect them to be in the range $10^{-2}$ to $10^{-3}$.
In Table 1 we give the values for two reference
models: the MSM with a 1 TeV Higgs $^6)$ 
as well as for a QCD-like model $^7)$
\begin{table}	
\begin{center}
\begin{tabular}{|l|cc|}\hline
& $L_1$ & $L_2$\\ \hline
MSM  ($M_H\sim1$ TeV)& 0.007 & -0.002 \\
QCD-like & -0.001 & 0.001 \\ \hline
\end{tabular}
\vspace{-.1cm}
\caption{\small Chiral Parameters for different reference models.}
\end{center}
\end{table}

We therefore have an answer to the first question. In case the SBS is strongly
interacting, we can measure the chiral parameters $L_1$ and $L_2$.
We will now review a study of the 
LHC sensitivity to the chiral parameters in the hardest non-resonant case $^8)$.

 In Table 2 we give the statistical significance (s=signal/$\sqrt{\hbox{backg.}}$)
to distinguish between the "zero model" (where all the couplings are
set to zero) and a model with some given values of $L_i$.
Following $^8)$, we give results for 400fb$^{-1}$ of
integrated luminosity at $\sqrt{14} \mbox{TeV}$. 
That corresponds to both experiments
working at full design luminosity during two years.
We only consider the processes $W^\pm Z^0\longrightarrow W^\pm Z^0$
and $W^+W^-\longrightarrow Z^0 Z^0$ in
 the cleanest decays, where the $W$'s and the $Z$'s
decay to $\nu_ee,\nu_\mu\mu$ and $e^-e^+,\mu^-\mu^+$, 
respectively.  For further details on the calculation we refer to  \cite{CMS}.
Notice that we are giving statistical significances with and without "jet tagging",
that could help us to separate $V_LV_L$ scattering from other processes involving quarks. We will comment the results in the conclusions.

\begin{table}  
\begin{center}
\begin{tabular}{||l||c|c|c|c||c|c|c|c||}   \hline \hline
&  \multicolumn{4}{c||}{$L_1$} &  \multicolumn{4}{c||}{$L_2$}\\  \cline{2-9}
 & $10^{-2}$ & -$10^{-2}$ &  $5\times10^{-3}$ & 
-$5\times10^{-3}$ & $10^{-2}$ & -$10^{-2}$ &  $5\times10^{-3}$ & 
-$5\times10^{-3}$\\  \hline 
${s|}_{W^{\pm}Z^0}$ & 1.4 & {\bf 5.2} & 1.2 & 2.0  & 1.4 & {\bf 9.6} & 0.4 & 3.4 \\  
${s|}_{Z^0Z^0}$ & {\bf 7.6} & 1.8 & 2.4 & 0.2 & {\bf 3.8} & 1.8 & 1.0 & $\simeq$0 \\  \hline
${s|}_{W^{\pm}Z^0\; tagging}$ & 2.0 & {\bf 8.4} & 1.8 & {\bf 3.4} & 2.0 & {\bf 15} & 0.6 & {\bf 5.4} \\  
${s|}_{Z^0Z^0\; tagging}$ & {\bf 13.2} & {\bf 3.6} & {\bf 4.6} & 0.4 & {\bf 7} & {\bf 3.6} & 1.8 & 0.2 \\ \hline
\hline 
\end{tabular}
\caption{Statistical significances
for different values of $L_1$ and $L_2$ at LHC.}
\end{center}
\end{table}


{\bf Resonance spectrum}

\vskip .3cm

Resonances and the saturation of unitarity
are the most characteristic features
of strong interactions.
In our case, we expect them to appear at 
the TeV scale. For instance, the MSM
becomes strong when $M_H\simeq1\mbox{TeV}$. In such case we expect a 
very broad scalar resonance around 1 TeV. In QCD-like models one 
expects a vector resonance around 2 TeV. 

Chiral lagrangians by themselves are not able to reproduce resonances.
Their amplitudes are obtained as polynomials in the momenta and masses,
and therefore they do not even satisfy unitarity. There is however a 
technique, known as the Inverse Amplitude Method (IAM), that is able to unitarize
these amplitudes. When applied to pion and kaon physics, it has successfully 
reproduced the lowest resonances $^9)$.
In the electroweak context it has been applied to the reference models
$^{10}$. Using the parameters in Table 1, a Higgs-like or a technirho are found in
their corresponding models.

What we present in Figure 1 is a scan of the $L_1,L_2$ parameter space using
the IAM. Using this method it is possible to obtain an estimate of the mass and
width of the resonances that would appear at LHC (below 3 TeV), depending
on the values of the chiral parameters $^{11})$.
We can find three types of resonances:

\hspace{.5cm} $\bullet$ Neutral and scalar. The usual Higgs boson is the typical example, that is why we will denote 
them as  $H$. If the width becomes
larger than 25\% of the mass we will call them $S_0$ (saturation). 

\hspace{.5cm} $\bullet$ Charged and vector-like. For example the technirho that appears in QCD-like models.
Generically we will denote them by $\rho$. Again, if they become too wide we will call them
$S_1$.

\hspace{.5cm} $\bullet$ Doubly charged and scalar. There are only models where
they are very light. In this channel, some values of the chiral parameters
are forbidden due to conflicts with renormalizability $^{12})$ and causality $^{11})$.
Such parameters correspond to the black area. Only very broad saturation shapes are admitted, but in any case they should be interpreted as resonances.

\vskip .2cm

{\bf Conclusions}

\vskip .2cm

$\bullet$ The study of this kind of physics will require the ultimate performance
at LHC ($\sqrt{s}$, and integrated luminosity), as well as in the detectors 
(jet tagging efficiency,etc).

$\bullet$ It seems possible to determine $L_1$ and $L_2$ to the 3$\sigma$ level
if their absolute value is not smaller than $10^{-3}$ (See Table 2). A precision of $10^{-3}$ seems 
extremely hard to achieve.

$\bullet$ Depending on the values of $L_1$ and $L_2$ we could find one or two resonances, a resonant channel and another one with saturation, or two channels with saturation. (See Figure 1)

$\bullet$ It is even possible that we do not see any resonance at all at LHC (grey area in Figure 1). 
The corresponding values of the $L_i$ are $\simeq10^{-3}$ and thus it could
be possible that we do not get any clear signal of strong interactions at LHC.

\renewcommand{\baselinestretch}{0.5}
\small\normalsize

\parbox{8.4cm}{


\mbox{\epsfysize=9.cm\epsffile{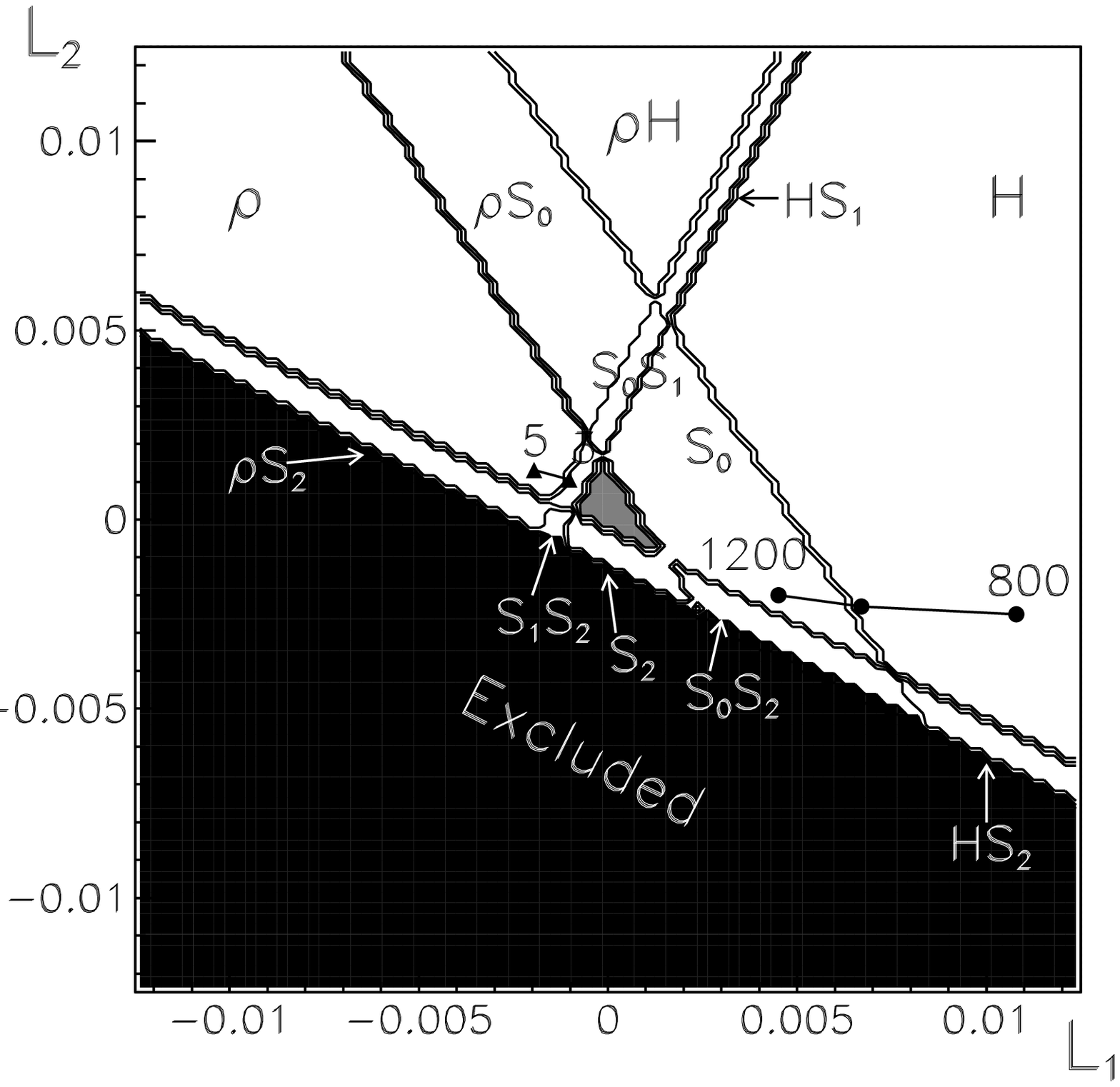}}

\vskip -.3cm

{\renewcommand{\baselinestretch}{0.3}
\small
{\bf Figure 1.-} 
The different areas in  parameter space
represent what resonances or saturation effects appear 
for different $L_1,L_2$ parameters.
There could be neutral scalar ($H$) or charged vector ($\rho$)
resonances, as well as saturation effects in the neutral scalar ($S_0$),
charged vector ($S_1$) or doubly charged scalar channel ($S_2$).}
}
\hfill
\parbox{8.4cm}{
{\bf References}
\small
\vskip .2cm

1. P.Sikivie et al. \NP{B173} (1980) 189.

2. M.S.Chanowitz, M.Golden and H.Georgi,
{\em Phys.Rev.} {\bf D36} (1987) 1490.

3. A.Dobado and J.R.Pel\'aez, {\em Nucl. Phys.}
 {\bf B425} (1994) 110.

4. A.Dobado and M.J.Herrero, {\em Phys. Lett.} {\bf B228}
 (1989) 495 and {\bf B233} (1989) 505;
 J.Donoghue and C.Ramirez, {\em Phys. Lett.} {\bf B234} (1990) 361.

5. S.Weinberg, {\em Physica} {\bf 96A} (1979) 327;
 J.Gasser and H.Leutwyler, {\em Ann. of Phys.} {\bf 158}
 (1984) 142; {\em Nucl. Phys.} {\bf B250} (1985) 465 and 517.

6. M.J.Herrero and E.Ruiz Morales,
{\em Nucl.Phys.} {\bf B418} (1994) 431;
{\em Nucl.Phys.} {\bf B437} (1995) 319.

7. A.Dobado and J.R.Pel\'aez, hep-ph/9604416.

8. CMS Technical Proposal. CERN/LHC94-38. LHCC/P1. (1994);
A.Dobado, M.J.Herrero, J.R.Pel\'aez,
E.Ruiz Morales and M.T.Urdiales \PL{B352} (1995) 400;
A.Dobado and M.T.Urdiales \ZP{B71} (1996) 659.

9. Tran N. Truong, {\em Phys. Rev. Lett.} {\bf 61} (1988) 2526,
 ibid {\bf D67} (1991) 2260. A.Dobado, M.J.Herrero and T.N.Truong, \PL
{B235} (1990) 134; A.Dobado and J.R.Pel\'aez, {\em Phys. Rev.} {\bf D47} 
(1992) 4883.

10. A.Dobado, M.J.Herrero and T.N.Truong, \PL
{B235} (1990) 129;A.Dobado, M.J.Herrero and J. Terr\'on, \ZP
{C50} (1991) 205; ibid 465

11. J.R.Pel\'aez, \PR{d55}(1997) 4193.

12. H.Georgi and M.Machacek, \NP{B262} (1985) 463.
R.S.Chivukula, M.J.Dugan and M.Golden, \PL{B336}(1994) 62.

}

\end{document}